# Finites elements and signal processing analysis of eccentricity and demagnetization faults in PMSM drivetrains: Approach for diagnosis

Naourez BEN HADJ*, Manel KRICHEN, Rafik NEJI

[1]*Department of Electrical Engineering, LETI-laboratory,
National School of Engineers of Sfax, Sfax University, Tunisia*
naourez.benhadj@enis.tn, krichenmanel2@gmail.com, rafik.nejii@gmail.com

Today, interest in automotive applications notably Hybrid Electric Vehicles (HEV) has risen due to environmental concerns and the modern society's energetic dependence. Consequently, it is necessary to study and implement in these vehicle fault diagnosis systems which allow them to be more reliable and safe enhancing its sustainability. Several research works have been focused on Fault Diagnosis approaches for electric machines due to it's influence in the automotive system availability. Our work fits into this context, firstly with a review of faults and diagnosis methods of the permanent magnet synchronous motor (PMSM) and secondly by using analytical-numerical approach based on finite element method and signal processing (FEMM-MATLAB coupling) in order to detect and analysis of eccentricity and partial demagnetization faults for surface mounted PMSM. So, this work mainly includes fault performance, harmonic component and signal spectrum characteristics, time-frequency analysis technique. The outcomes obtained are further compared with the motor characteristics under healthy operation to have distinct detection to faults.

*Keywords*: Permanent magnet synchronous motor; Finite element method; fault detection; Eccentricity; Demagnetization; Signal processing;

## 1. Introduction

Over the past decades, concerns about the safety and reliability of automotive applications have increased a lot due both to adopted regulation concerning the $CO_2$ emission limitations and to energetic dependence of the modern society. In this context, the orientation of the research was focused on transportation problems notably HEVs by treating the problems of electrical vehicles simulation, the new technologies for energy storage, the drivetrains design, integration and control.

However, the complexity of the drive-trains in HEV makes them fragile in terms of availability. So, in order to be competitive on the market, the reliability of the electrical drive-train is considered by the searchers as one of the most important obstacles against a large public use of the HEV. This work focus on the problem of availability of HEV drivetrains based on PMSM fault diagnosis since the requirements of security, availability and reliability of industrial processes have become increasingly more severe. Indeed, the PMSM are increasingly integrated into various HEVs as a traction motor or for ensuring an auxiliary function as the steering assistance. During the last ten years, this success was confirmed due to multiple benefits offered by these motors namely a high torque and power density, simple and compact rotor structure, , lower losses, better dynamic performance, easy maintenance, and very good reliability [1]. Given the above-mentioned advantages, the production of PMSMs faces some hurdles, such as possible PM demagnetization, a higher price due to the use of expensive permanent magnets (PMs), operating temperature sensitivity, and the requisite position sensors for their control [2]. In this context the PMSM will evolve to meet the socioeconomic constraints notably the high degree of availability (continuity of service) despite the very severe operating conditions. To meet this societal demand, a good knowledge of the behavior of these machines in extreme environment, or under mechanical, electrical and magnetic faults conditions is more essential than ever (vibrations, wide speed ranges and temperatures, repetitive starts and stops, eccentricity, demagnetization PM, voltage fluctuation ...).

---

* Typeset names in 8 pt Times Roman. Use the footnote to indicate the present or permanent address of the author.

In this paper, we start with an overview of faults and faults diagnosis methods in PMSM propulsion systems. According to consideration and analysis different diagnosis methods presented by literatures, Fast Fourier transform (FFT) has been selected as a proper method in stationary state operation. Whereas, section 3 describes the proposed Methodology of surface mounted permanent magnet synchronous motor (SPMSM) analytical modeling, analysis by applying finite element method. Furthermore, two types of faults viz eccentricity and demagnetization faults in SPMSM has been studied. Analytical model and simulation model have been developed for characterizing the eccentricity faults of this motor. Also, a study on the SPMSM partial demagnetization faults has been performed and explained. Section 4 presents the signal processing analysis for SPMSM eccentricity and demagnetization faults. This allows larger diagnosis knowledge about the SPMSMs. Section 5 concludes the work.

## 2. Signal-based fault diagnosis methods in PMSM propulsion systems

Using PMSM has high reliability in HEVs but some damage is sustained under conditions of continuous operation and under continuous load and speed changes. Faults in PMSMs are divided into three types: Electrical faults (ELF), magnetic faults (MAF) and mechanical faults (MEF). ELF include short-circuit stator phase windings, the open-circuit of the entire phase, the incorrect attachment of the motor windings, and the grounding faults. The stator windings inter-turn short circuit is the most common fault in PMSM [3,4,5]. MEF include mainly magnet damage, shaft bending, bolt loosening, bearing faults, and eccentricity of the air gap. Bearing failures, which lead to approximately 40–50 per cent of all motor faults [6,7]. Faults of eccentricity include three types which are dynamic eccentricity (DE), static eccentricity (SE), and mixed eccentricity (ME). Faults in eccentricity can cause noise, ripple torque, and additional vibration. And when the eccentricity fault is severe, it can create friction between the stator and the rotor, and damage the stator or rotor core, a performing the motor's normal operation [9,10,11].

PMs in the PMSM can be demagnetized by large short-circuit currents from inverter or stator faults, magnet harm, large stator currents, high temperature, and magnet aging itself. Such resistance will slowly induce demagnetization of the PM over the course of long- activity. When this demagnetization occurs on part of magnetic pole, it is demagnetized partially, and complete demagnetization corresponds to the entire magnetic pole. According to several studies, the torque would be inadequate when the PMSM-specific demagnetization fault occurs due to flux contact ripples, allowing the current increase to provide the necessary torque, which will boost the temperature and intensify the demagnetization in effect. At the same time the torque fluctuation can also result in irregular vibration and acoustic noise. We can lower motor output and efficiency and have a negative effect on the PMSM's normal activity [12,13].

Three motor fault diagnosis methods are used commonly viz signal-based fault diagnosis methods (SFDM), model-based fault diagnosis methods (MFDM) and Data-driven intelligent diagnosis methods (DDIDM). Most generally, SFDM are used to process and remove various features from the motor signals, and these features are then identified manually using experience or knowledge-based approaches. The MFDM can determine the fault type. After different faults occur in the motor, the MFDM can estimate the signal performance by comparing the output of the motor model with the actual output data. DDIDM can automatically evaluate the fault state based on expert knowledge but, thanks to work in artificial intelligence and machine learning, data-driven smart diagnostic systems which do not depend on prior knowledge, have recently demonstrated large application prospects [15,8,16].

Some complicated fault phenomena may appear when faults occur in motor. There are non-electrical quantities like vibration, sound, heat, gas, light, radiation and also electrical quantities like current, voltage, power, frequency and so on. From the signals of current, vibration and so on, these methods can extract fault features and identify the performance of faults [17,8]. Some research teams who have combined different signals to build a fault diagnosis system of PMSM. The motor current signal analysis (MCSA) has been more studied [18,8]. Generally, Time domain methods, frequency domain methods, and time-frequency analysis methods are the three types of signal processing methods. The FFT is the classical method of frequency domain analysis. The signal in Fourier transform is represented as a superposition of several cosine or sine functions. This demonstrate clearly the frequency distribution of the signal, where the amplitude and frequency of harmonic components can be used as the features of different faults [23].

The most noticeable feature among short circuited winding fault is the increase of the third harmonic amplitude [8,30,31]. Mechanical faults cause distortions in the flux distribution within the engine, which in effect lead to some current harmonics in the stator current [14,16,8]. However, the distributed MMF will not be sinusoidal in case of demagnetization fault. So, the failed and normal portions of the MMF will produce current with multiple frequencies, which means, low frequency components will appear near the fundamental wave in signal of the current [19]. Time information is lost with the frequency domain approaches, and it is often difficult to differentiate with precision similar harmonics. An improved method based on FFT is used which is short-time Fourier transform (STFT) [20]. By dividing the signal into small time windows, It realizes time-frequency analysis. In [8], Rosero et al. mixed the STFT and Gabor spectra to analyze the PMSM current signal since, STFT performs the worst in several time-frequency analysis techniques.

Discrete wavelet transform (DWT), continuous wavelet transform (CWT) are two categories of Wavelet Transform (WT). With WT, the signal can be decomposed into signals with different resolutions at different frequency bands, making it suitable for accurate and flexible fault detection of PMSM. To identify the short circuit winding phase in PMSM, Obeid et al. have used the WT methods [8]. Rosero et al. have focused on mechanical faults and have evaluated the degree of demagnetization of PMSM by calculating the energy in the detail of DWT [21]. To extract the characteristics of the PMSM inter-turn short circuit fault, An-ping et al. have used the wavelet packet band energy analysis method. This method is to accurately divide the frequency band of the current signal by wavelet packet transform (WPT) and consider the signal energy of the appropriate frequency bands as the feature vector of the fault. Finally, it is also impossible to have high accuracy in both time and frequency at a certain frequency band like STFT [8].

The Hilbert-Huang transform (HHT) combines empirical mode decomposition (EMD) with Hilbert transform. it is considered a time-frequency analysis technique. By EMD, the signal is first decomposed into multiple internal mode functions (IMF). Then, to calculate the instantaneous frequency of the original signal from the IMF, the Hilbert transform is used. This approach resolved the shortcomings of STFT and WT, which is their time-frequency resolution which is closely related to the selection of window size or window function. Using HHT, Urresty et al. have measured the PMSM current Hilbert energy spectrum to detect the short-circuit stator fault. Furthermore, due to its sensitivity to the transient frequencies, HHT is ideal for dynamic signal analysis. also, HHT was applied to diagnose the demagnetization fault, and the results showed that there were still good results in dynamic operating conditions and high-speed [8].

The Wigner-Ville distribution (WVD) is a time frequency energy density. The joint time and frequency resolution provided by the WVD is the best better than other time-frequency analysis methods and its calculation cost is not high [22,8]. Nevertheless, if the signal contains more than one frequency, cross-term interference will result from WVD, which is undoubtedly pretty unfavorable for fault diagnosis. Some researchers have combined WVD with EMD, included in HHT, to analyze the current by analyzing the demodulated IMF with WVD [23]. Rosero et al. used a smoothed pseudo Wigner Ville distribution (SPWVD) and Zao-Atlas-Marks distribution (ZAM), which can eliminate interference in WVD-based PMSM fault diagnosis [24,8]. Even so, its implementation on PMSM remains to be further studied.

## 3. PMSM Modeling and analysis under healthy and faulty states

### 3.1. *Proposed Methodology*

As described in figure 1, the proposed methodology for the SPMSM modeling and analysis is founded on analytical modeling, finite element methods (FEM) analysis and signal based faults analysis.

The SPMSM analyzed in this paper has a power of 62 kW, four pairs of poles and six main teeth, an interposed tooth is inserted between the two main teeth to boost the wave shape and to reduce the leakage risk. Each phase winding consists of two diametrically opposed coils with concentrated windings which means that the coil circles are wound directly around a stator tooth. In this study, the air gap is 2mm and the height of the PM is 7.76 mm. The properties of the SPMSM are presented in table 1. The 2D field analysis is performed using the FEMM 2D program. The properties of used permanent magnets are the remanence $B_r$=1.16T, the coercivity $H_c$=-895kA/m and the permeability of the recoil, $\mu_r$=1.044.

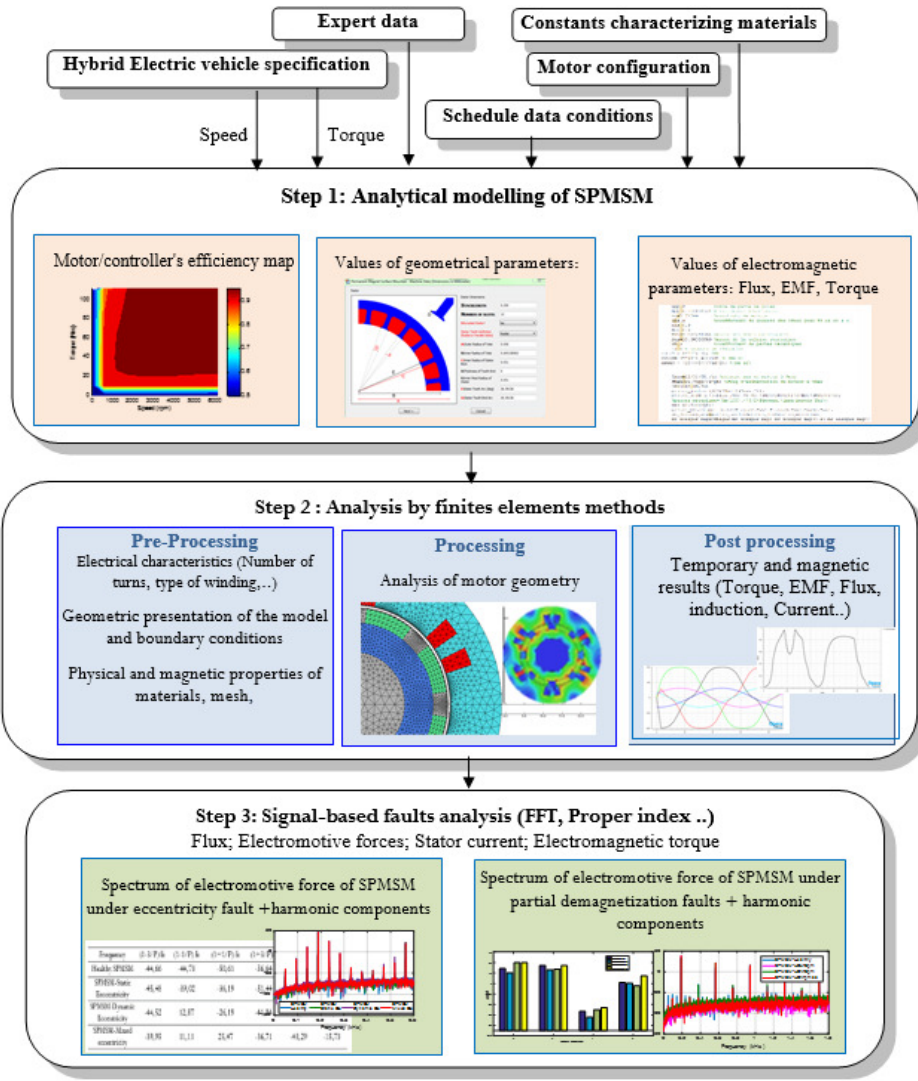

Fig. 1. Schematic of modeling and analysis of SPMSM

The magnetic induction of the analyzed SPMSM at no load and the field lines distribution for a rotor position are shown respectively in Figure 2. (a) and (b). The proposed motor was simulated at first time under eccentricity fault and at second time under partial demagnetization fault.

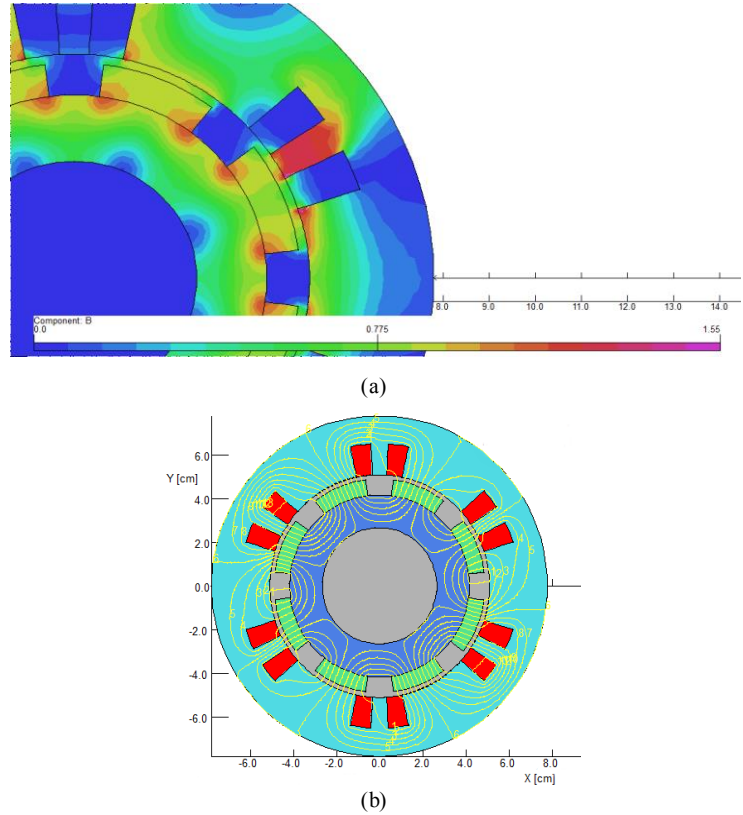

(a)

(b)

Fig.2. (a) Magnetic induction in SPMSM at no load, (b) Cross section with field lines distribution

Table 1. Properties of the 62 kW SPMSM used in this study.

| **Parameters** | **Value** | **Parameters** | **Value** |
|---|---|---|---|
| Stator exterior diameter | 174 mm | Magnets number | 8 |
| Stator interior diameter | 102 mm | Slots number | 12 |
| rotor exterior diameter | 82 mm | Poles number | 8 |
| Core stack length | 250 mm | Terminal current | 59.5 A |
| Magnet height | 7.76 mm | Supply frequency | 188.3 Hz |
| Core stack length | 250 mm | Rotational speed | 2824 rpm |

After applying finite elements study in healthy mode, the obtained flux is equal to 0.0055 Wb which validates the analytical model which gives a flux value equal to 0.00589 Wb (figure 4.a). Too, the Electromotive force obtained by the FEM, is equal to 652V which validates the analytical model (697 V) and finally the electromagnetic torque obtained by the FEM, is equal to 217 Nm which validates the analytical model well since the torque oscillates around the mean value found analytically which is 210 Nm (figures 5.a and 6.a).

### 3.2. *Analytical computation of SPMSM*

The angular speed $W_m$, the torque $T_m$, and the power $P_m$ required of the SPMSM to move the vehicle are expressed by eq. (1), (2), and (3), respectively [25]. Where *V* is the rated vehicle speed, $K_r$ is the gear ratio and $R_{wh}$ is the wheel radius.

$$W_m = \frac{V}{R_{wh}} k_r \quad (1)$$

$$T_m = \frac{FR_{wh}}{k_r} \quad (2)$$

$$P_m = T_m W_m \quad (3)$$

Where F is the vehicle propulsion force defined as follow. Where m is the vehicle mass, g is the vehicle gravity, *Crr is the r*olling resistance coefficient, *ρ is the v*ehicle air density, *$C_D$ is the c*oefficient of aerodynamic drag, $A_f$ is the vehicle frontal area, *Θ is the g*rade and *a is the* Vehicle acceleration.

$$F = mgC_{rr} + \frac{1}{2}\rho C_D A_f V^2 + ma + mg\sin(\theta) \quad (4)$$

In stator geometrical sizes of the SPMSM, the slot average width $L_{enc}$ is expressed by (5), where *$D_m$ is the* average diameter of the SPMSM, *e* is the air gap thickness, $H_d$ is the tooth height and *$A_{enc}$* is the slot angular width.

$$L_{enc} = \frac{D_m + e + H_d}{2} A_{enc} \quad (5)$$

The principal tooth section $S_d$ and the inserted tooth section $S_{di}$ are expressed as follow. Where *$L_m$ is the m*otor length, $A_{dent}$ is the principal tooth angular width and $A_{denti}$ *is the i*nserted tooth angular width.

$$S_d = \frac{D_m + e}{2} A_{dent} L_m \quad (6)$$

$$S_{di} = \frac{D_m + e}{2} A_{denti} L_m \quad (7)$$

The slot section $S_e$ is expressed by:

$$S_e = A_{enc}\frac{D_m + e}{2}L_m = \frac{1}{2}\left[\frac{2\pi}{Nd} - A_{dent} - A_{denti}\right]\frac{D_m + e}{2}L_m \quad (8)$$

In the rotor geometrical sizes of the SPMSM, the expression of the magnet height $H_a$ is obtained by the application of the Ampere theorem.

$$H_a = \frac{\mu_a B_e e}{M(T_a) - \frac{B_e}{K}} \tag{9}$$

The remanent induction of the magnet $M(T_a)$ at $T_a°C$ is defined by :

$$M(T_a) = M[1 + \alpha_m(T_a - 20)] \tag{10}$$

The rotor yoke thickness $H_{cr}$ is defined:

$$\frac{\emptyset}{2} = B_{cs} H_{cr} L_m K_{fu} = \frac{B_e S_d}{2} \implies H_{cr} = \frac{B_e S_d}{2 B_{cs} L_m K_{fu}} \tag{11}$$

Where $N_d$ is the number of principal tooth, $\mu_a$ is the relative permeability of magnets, $B_e$ is the magnetic induction in airgap, $\alpha_m$ *is the t*emperature coefficient, *$B_{cs}$ is the m*agnetic induction in stator yoke and $T_a$ is the magnet temperature.

### 3.2. *Finite element study and analysis for faults in SPMSM*

#### 3.2.1. *Eccentricity faut*

In healthy motor, the rotation center of the rotor is identical to the stator geometric center. In SE, the rotation center $O_W$ is identical to the rotor center $O_r$, Static eccentricity ratio $\delta_s$ and the airgap width $g_s$ are expressed respectively by eq. 12 and 13. Figure 3.a, illustrates the distance between the stator center and any point M on surface of the rotor. Also, it shows that $\varphi_s$ is equal to $\theta_s$ which is the reference angle of the stator [27]. In DE, the rotation center is identical to the stator center $O_S$, the dynamin excentricity ratio $\delta_d$ and the airgap width $g_d$ are expressed respectively by eq. 14 and 15 [27]. Figure 3.b, illustrates the PMSM geometry with DE. Also, from the reference point of the stator frame, it is obvious that the width of the airgap at a fixed stator angle depends on electrical rotational speed $\omega_S$ and the pole pairs number p [28]. In mixed eccentricity, the rotation center, the rotor center are different to the stator center. The mixed excentricity ratio $\delta_m$ and the airgap width are expressed respectively by eq. 16 and 17 [26,27]. Figure 3.c depicts the motor geometry with ME. It indicates the airgap broadness in the PMSM depends on the rotor's mechanical angle.

$$\delta_s = \frac{|O_s O_w|}{g} \tag{12}$$

$$g_s(\theta) = g(1 - \delta_s \cos(\theta)) \tag{13}$$

$$\delta_d = \frac{|O_r O_w|}{g} \tag{14}$$

$$g_d = g(1 - \delta_d \cos(\frac{\omega_s}{p} t - \theta_s)) \tag{15}$$

$$\delta_m = \delta_s + \delta_d = \frac{|O_s O_r|}{g} = \sqrt{\delta_s^2 + \delta_d^2 + 2\delta_s \delta_d \cos(\theta)} \tag{16}$$

$$g_m(t) = R_s - \delta_m g \cos\left(\frac{\omega_s t}{p} - \varphi_m\right) - \sqrt{R_r^2 - \delta_m^2 g^2 \sin^2\left(\frac{\omega_s t}{p} - \varphi_m\right)} \tag{17}$$

Where: g is the uniform airgap

θ is the angular position around the airgap from 0° to 360°

θs is the fixed stator angle in dynamic eccentricity

$R_r$ is the rotor radius and $R_s$ is the stator radius

$\varphi_m$ is the Mixed eccentricity transfer angle

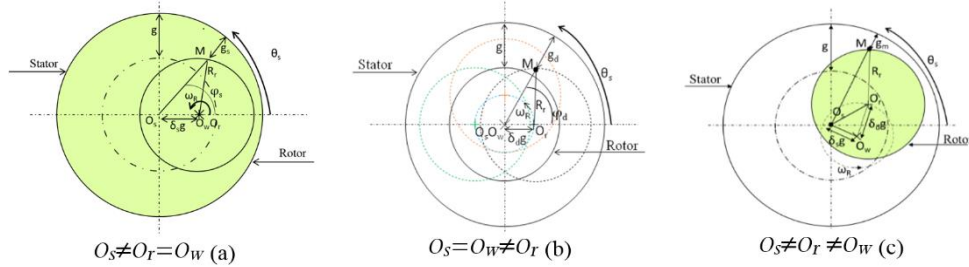

Fig 3. PMSM geometry (a) with SE (b) with DE (c) with ME

The flux linkage presented in Figures 4.1.a, 4.1.b, 4.1.c is from four cases which are the healthy case, the case under 40% of static eccentricity, the case under 40% of dynamic eccentricity and the case under mixed eccentricity (40% of static and 40% of dynamic eccentricities). In different eccentricity fault, a non-uniform magnetic flux density develops around the rotor, causing perturbations in the flux. The SE has a very slight influence on the shape and on the flux amplitude for the healthy SPMSM case (the amplitude flux variation is equal to 0.69%). Also, the amplitude flux variation in the case of DE is around 5.74% compared to that of the healthy case. In the ME case, the flux amplitude varies more significantly in relation to that in the case of static and dynamic eccentricity. (Amplitude flux variation is equal to 7.19% compared to the healthy case). The back-EMF per phase in the studied SPMSM, is calculated in all coils with the phase winding connected in parallel. In each phase, the back-EMF $E_{phase,i}$ in $i^{th}$ coil is expressed by eq. 18 where N is the turns number per phase and $\varphi_{phase,i}$ is the flux linkage of the $i^{th}$ coil [28]. The back-EMF per phase presented in Figures 5.1.a, 5.1.b, 5.1.c is from the four cases which are already defined before. It is shown from these figures that the EMF magnitude varies depending on the width of the airgap and this variation is more significantly in case of ME. The SPMSM electromagnetic torque for the healthy case and the three type of eccentricity is illustrated in Figures 6.1.a, 6.1.b, 6.1.c. An influence can be visualized in the PMSM torque compared to the healthy case [29]. This is because of the sensitivity of the currents and the EMFs used for evaluating the electromagnetic torque by eq. 19. The peak-to-peak torque ripple is equal to 15.17% in healthy case. It increases to 23.08% in SE, to 37.27% in DE and to 40.48% to ME.

$$E_{phase,i} = -N \frac{d\varphi_{phase,i}}{dt} \tag{18}$$

$$C_{em}(t) = \frac{E_{phase1}(t) \times I_{phase1}(t) + E_{phase2}(t) \times I_{phase2}(t) + E_{phase3}(t) \times I_{phase3}(t)}{\Omega(t)} \quad (19)$$

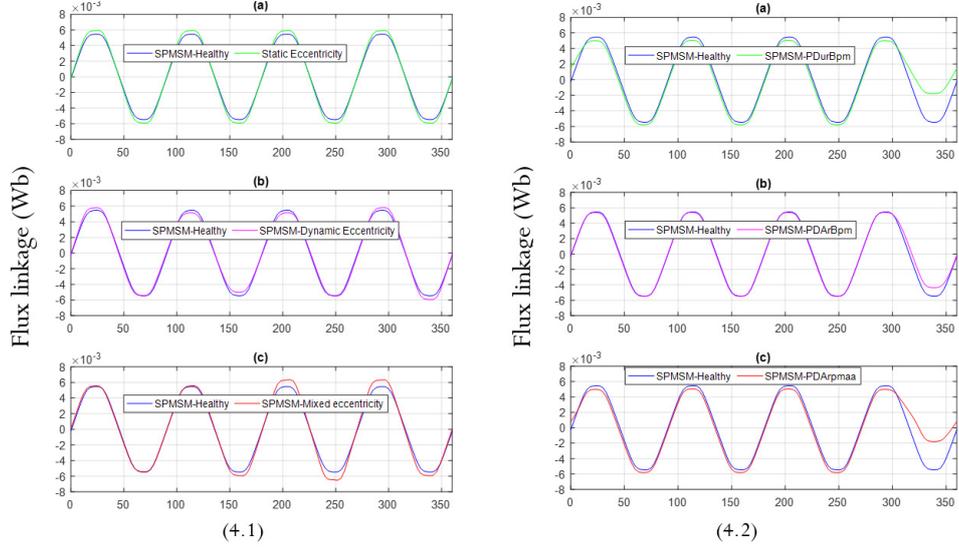

Fig. 4. Flux linkage in SPMSM (4.1.a) under healthy and 40% of SE (4.1.b) under healthy and 40% of DE (4.1.c) under healthy and 40% of ME (4.2.a) under healthy and PD$_{urBpm}$ (4.2.b) under healthy and PD$_{ArBpm}$ (4.2.c) under healthy and PD$_{Arpmaa}$

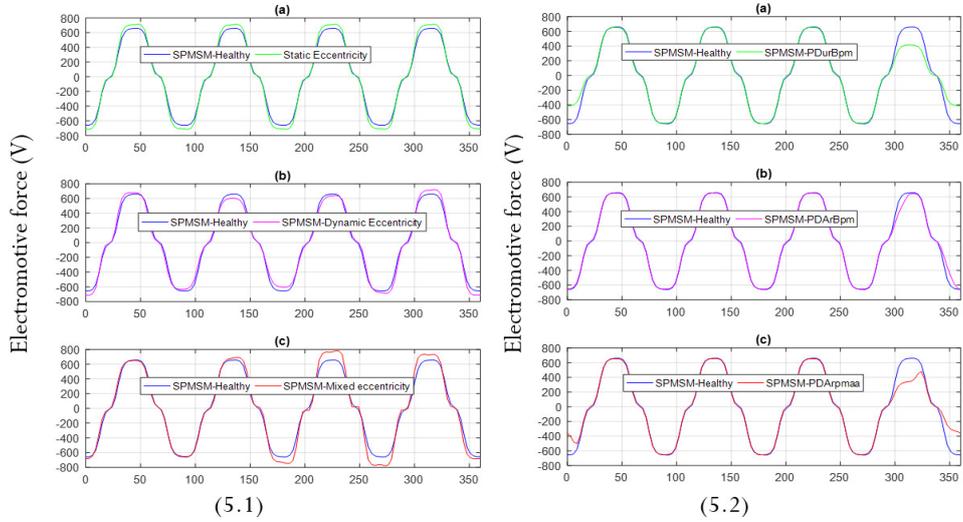

Fig. 5. Electromotive force of SPMSM (5.1.a) 40% of SE (5.1.b) 40% of DE (5.1.c) 40% of ME (5.2.a) under PD$_{urBpm}$ (5.2.b) under PD$_{ArBpm}$ (5.2.c) under PD$_{Arpmaa}$

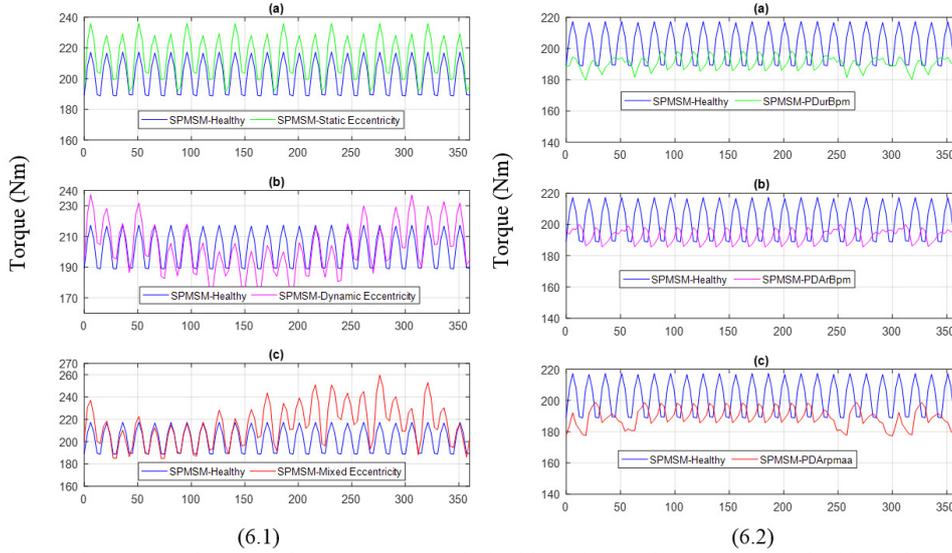

(6.1) (6.2)

Fig. 6. Electromagnetic torque of SPMSM (6.1a) under healthy and 40% of SE (6.1.b) under healthy and 40% of DE (6.1.c) under healthy and 40% of ME (6.2.a) under healthy and $PD_{urBpm}$ (6.2.b) under healthy and $PD_{ArBpm}$ (6.2.c) under healthy and $PD_{Arpmaa}$.

### 3.2.2. *Demagnetization faut*

In this section, we look about the partial demagnetization faults which are manifested in three fault scenarios viz the partial reduction of a single $B_{pm}$ Magnet Fault ($PD_{urBpm}$), asymmetric reduction of a single $B_{pm}$ Magnet fault ($PD_{ArBpm}$) and asymmetric reduction of a single PM arc angle fault ($PD_{Arpmaa}$) [27,32].

In the two cases which are $PD_{urBpm}$, $PD_{ArBpm}$, the material for the chosen magnets was replaced with a material that has the same conductivity and relative permeability as the healthy magnets but a lower magnetic flux density PM compared to the healthy magnet. A reduction of 75% of the healthy $B_{pm}$ (M = 0.25) was investigated to represent partial demagnetization.

For the case of the partial demagnetization fault by asymmetric reduction of a single PM arc angle ($PD_{Arpmaa}$), the damage of the magnet was presented by changing the geometry of the Permanent magnet segment. A reduction of 75% of the healthy magnet arc angle (X = 0.25) was simulated to represent the reduction of the arc length magnet in the case of partial demagnetization.

As results of the three types of partial demagnetization with 75% fault severity in comparison with healthy condition, figures 4.2-5.2 depict respectively the flux, the EMFs as a function of rotor position and figure 6.2 presents the EM torque as a function of the rotor position. The demagnetized PMs produces less flux and destroys his sinusoidal form than healthy PMs (Figure 4.2), weakens the EMFs mostly the two cases viz the $PD_{urBpm}$ and the $PD_{Arpmaa}$ (Figure 5.2).With the cases of $PD_{urBpm}$ and $PD_{Arpmaa}$ arc angle have more influence than the case of faulty asymmetric reduction of a single $B_{PM}$ Magnet ($PD_{ArBpm}$). we note that the flux in three cases of partial demagnetization decreases compared to its value in healthy case (the variation is 08.36% in PD$_{UrBpm}$, 02.03% in PDArBpm and 08.3% in PD$_{ARpmaa}$).

Figure 6.2 gives the Electromagnetic torque (EM) as a function of the rotor position. This figure shows that the demagnetized PMs produces less torque than healthy PMs. We conclude that PD$_{Arpmaa}$ is the more serious fault for partial demagnetization cases.

## 4. Signal processing analysis applied for SPMSM under faulty states

To control and operate the SPMSM under healthy and faulty conditions, electromotive forces signals were obtained from simulation results for processing in the abc frame of reference. After that, Fast Fourier Transform was applied to the electromotive force signals to extract their spectra.

### 4.1. *Signal -based eccentricity fault results and analysis*

In this part of the paper, a novel frequency pattern is introduced for statistic, dynamic and mixed eccentricities, and to diagnose the different fault eccentricities, the amplitude of sideband components (ASBCs) at frequencies extracted from the frequency pattern is used. According to the research [8], the characteristic frequencies for PMSM eccentricity faults are expressed by eq. 20. Moreover, Ebrahimi et al., speak about the harmonic at $f_s(1 - \frac{3}{p})$ is more related to SE faults, while the harmonic at $f_s(1 + \frac{1}{p})$ is related to both SE and DE [21].

$$f_{ecc} = f_s(1 \pm \frac{2k-1}{p}) \tag{20}$$

Where: fs is the frequency of the power supply (equal to 188.3 Hz),

    k is a positive integer (*k* = 1, 2, 3…),

    p is the number of pole pairs.

Figures 7.1 illustrate the spectrums of the electromotive force under healthy, 40% SE, 40% DE and 40% ME, with considerable amplitudes in the frequencies 47.075Hz, 141.225Hz, 235.375Hz, 329.525Hz, 423.675Hz and 517.825Hz. The amplitude of sideband components of the pattern frequency mentioned above can be used for the detection of the different types of eccentricity. Table 2, reports the ASBCs at the frequencies already defined due to static, dynamic and mixed eccentricities. What can be noticed is that, in the case of ME fault, the amplitude of the sideband components was higher compared to that recorded in static and dynamic eccentricities faults. This has been proposed as a measure to differentiate the two types of eccentricity fault. To recap, the most dangerous fault is mixed eccentricity compared to static and dynamic eccentricity which reduces the machine performance. In addition, from figure 8, the SPMSM under mixed eccentricity is more affected than in the static and dynamic eccentricity. The frequency ((1+3/P)fs) and ((1+5/P)fs) are respectively more related to dynamic and static eccentricities. The frequency ((1-3/P)fs) is more related to mixed eccentricity.

### 4.2. *Signal -based demagnetization fault results and analysis*

To detect and distinct the different demagnetization fault, a frequency pattern is introduced, and the ASBC is used at frequencies extracted from the frequency pattern. Melicio.J and all define the magnetic flux density distribution as a function of rotor position $B(\theta_r)$ for healthy rotor by eq. 21. This airgap flux distribution is a periodic, odd function with half cycle symmetry containing odd harmonics components of the fundamental frequency in the stator current signal (eq. 22)[27].

$$B(\theta_r) = \sum_{n=1,3,5,\ldots}^{\infty}(B_{PM})_n \sin(np\,\theta_r) \tag{21}$$

With $\quad (B_{pm})_n = \frac{4B_{PM}}{\pi n}\left[\sin(\frac{n\pi}{2})\sin(\frac{np\alpha_M}{2})\right]$

$$f_{Healthy} = (2k-1)f_s \tag{22}$$

Where: $(B_{PM})_n$ is the nth order PM flux density distribution Fourier coefficient,

n is the harmonic order number ($n$ = 1, 3, 5…),
$\theta_r$ is the rotor angular position in mechanical degrees,
$\alpha_M$ is the magnet arc angle in mechanical degrees.
k is a positive integer ($k$ = 1, 2, 3…),

The harmonics of the PM field disturbance induced by the uniform demagnetization fault do not compromise the half-cycle symmetry or the odd periodic spatial distribution of the PM field. Thus, it only induces changes in the amplitude harmonic orders which should exist in the airgap of a healthy machine (eq. 23). For instance, it can be shown that the harmonic series in eq. 21 will be modified to eq. 24 when M is defined as the flux magnitude fault severity indice, which can take values from 0 in total demagnetization to 1 in healthy PM. The possible electromotive forces spectral frequencies for a uniform demagnetization fault, $f_{UD}$ will be identical to those existing in the healthy machine [32]. For partial demagnetization faults, such as PD$_{urBpm}$, PD$_{ArBpm}$, the field distribution property of half cycle symmetry is compromised. Therefore, $f_{PD}$ defines the possible electromotive force spectral frequencies in the presence of partial demagnetization faults [19].

$$f_{UD} = (2k-1)f_s \tag{23}$$

$$B_{UD1}(\theta_r) = \sum_{n=1,3,5,\ldots}^{\infty} \frac{4MB_{PM}}{\pi n}\left[\sin(\frac{n\pi}{2})\sin(\frac{np\alpha_M}{2})\right]\sin(np\,\theta_r) \tag{24}$$

$$f_{PD} = (1 \pm \frac{2k-1}{p})f_s \tag{25}$$

Table 2.2, reports the ASBCs at the frequencies already defined due to the three cases of partial demagnetization and in case of healthy SPMSM. Figure 7.2 shows for a bandwidth of 0–700 Hz, the simulation results for electromotive force spectrum for healthy SPMSM and for the three partial demagnetization considered in table 2. Figure 8, show that the partial

1414  *Author Names*

demagnetization of PM increases the amplitude of sideband components at all frequencies. The ASBCs for $PD_{Arpmaa}$ is more significative than other cases of partial demagnetization. The frequency $(1+7/p)$ fs is more significative to detect the partial demagnitisation types. The ASBCs for PDArpmaa is more than the ASBCs for $PD_{ArBpm}$ in all frequency already defined. The data for the studied SPMSM, show that the fault signature patterns caused by partial demagnetization fault types produce distinct features in the electromotive force signal which could be used to distinguish partial demagnetization defect types.

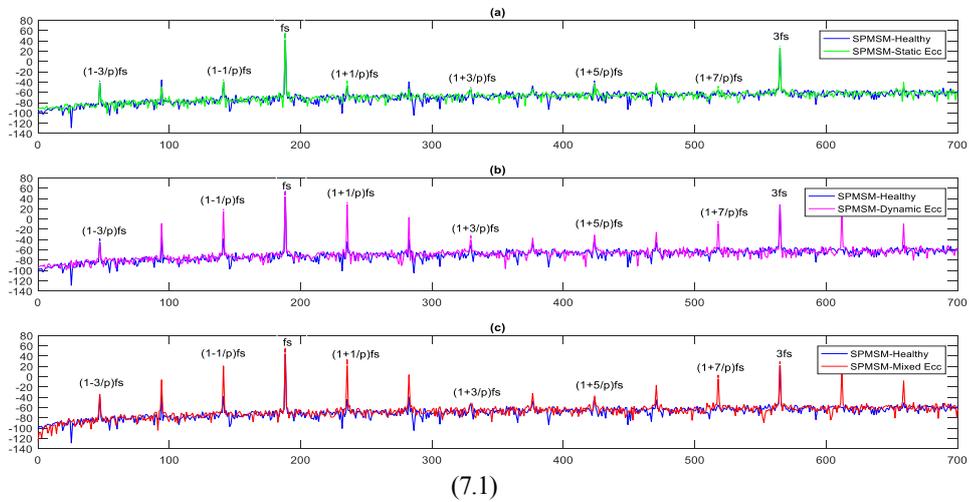

(7.1)

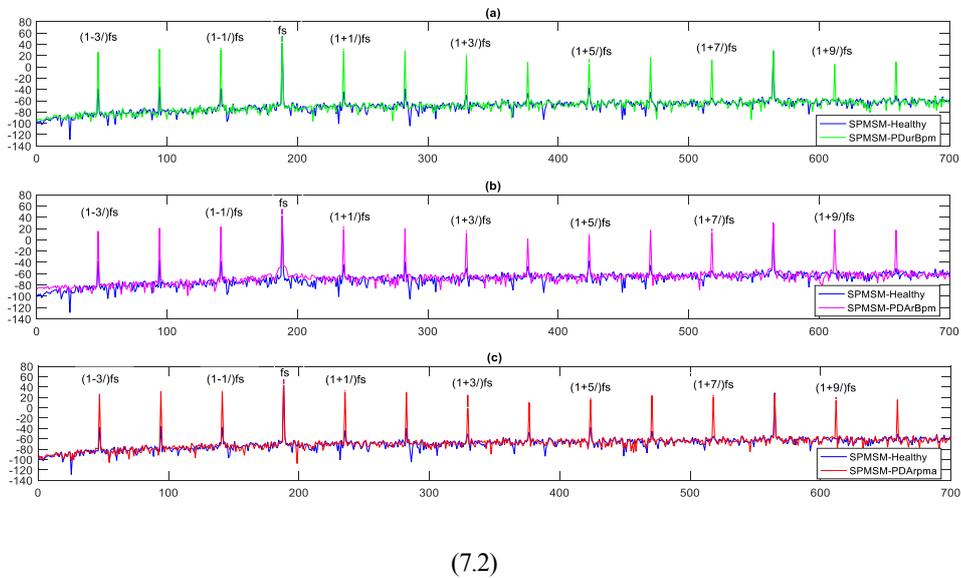

(7.2)

Fig.7. Spectrum of electromotive forces (7.1.a) under healthy and 40% of SE (7.1.b) under healthy and 40% of DE (7.1.c) under healthy and 40% of ME (7.2.a) under healthy and $PD_{urBpm}$ (7.2.b) under healthy and $PD_{ArBpm}$ (7.2.c) under healthy and $PD_{Arpmaa}$

TABLE 2. Electromotive forces harmonic components (In Decibels)

| (2.1) Comparison due to eccentricities faults (In Decibels) | | | | | | |
|---|---|---|---|---|---|---|
| Frequency | (1-3/P)fs | (1-1/P)fs | (1+1/P)fs | (1+3/P)fs | (1+5/P)fs | (1+7/P)fs |
| Values | 47.075 | 141.225 | 235.375 | 329.5 | 423.6 | 517.8 |
| Healthy SPMSM | -38.64 | -38.70 | -44.58 | -50.62 | -37.91 | -55.03 |
| SPMSM-Static Eccentricity | -41.44 | -35.41 | -37.76 | -51.50 | -43.92 | -48.28 |
| SPMSM-dynamic Eccentricity | -47.16 | 18.91 | 32.22 | -32.34 | -30.55 | -5.27 |
| SPMSM-Mixed eccentricity | -32.88 | 20.82 | 33.00 | -55.09 | -36.20 | 3.16 |
| (2.2) Comparison due to partial demagnetization faults (In Decibels) | | | | | | |
| Frequency | (1-3/P)fs | (1-1/P)fs | (1+1/P)fs | (1+3/P)fs | (1+5/P)fs | (1+7/P)fs |
| values | 47.075 | 141.225 | 235.375 | 329.5 | 423.6 | 517.8 |
| Healthy SPMSM | -38.64 | -38.71 | -44.58 | -50.62 | -37.91 | -55.03 |
| SPMSM-$PD_{urBpm}$ | 26.04 | 33.27 | 31.61 | 22.09 | 13.66 | 13.80 |
| SPMSM-$PD_{ArBpm}$ | 15.10 | 22.92 | 22.60 | 15.54 | 10.82 | 19.37 |
| SPMSM-$PD_{Arpmaa}$ | 26.08 | 33.63 | 32.67 | 24.35 | 18.49 | 24.14 |

The results in Figure 8 and detailed in table 2.2 show that, as predicted in (25), the existence of a partial demagnetization produces additional, even harmonics in the electromotive forces spectrum. To recap, the most risky fault is the partial demagnetization fault by asymmetric reduction of a single PM arc angle compared to other partial demagnetization faults which reduces the machine performance.

To summary, we note that partial demagnetization can be clearly identified for diagnosis by using harmonic components. We note also that partial demagnetization fault generate more harmonic component than eccentricity fault also the $PD_{Arpmaa}$ fault is more significative than other demagnetization types. The ME fault is more significative than other eccentricities types in except harmonic order 3 and 5. The partial demagnetization and eccentricity faults can be clearly identified for diagnosis by using harmonic components. Such defects produce asymmetry (there is an unbalance) and the harmonic component therefore exists in the SPMSM. Thus, the method of EMF signal processing is valid to detect and diagnose a partial demagnetization and eccentricity in the SPMSM since electromotive harmonics composition analysis have obvious results.

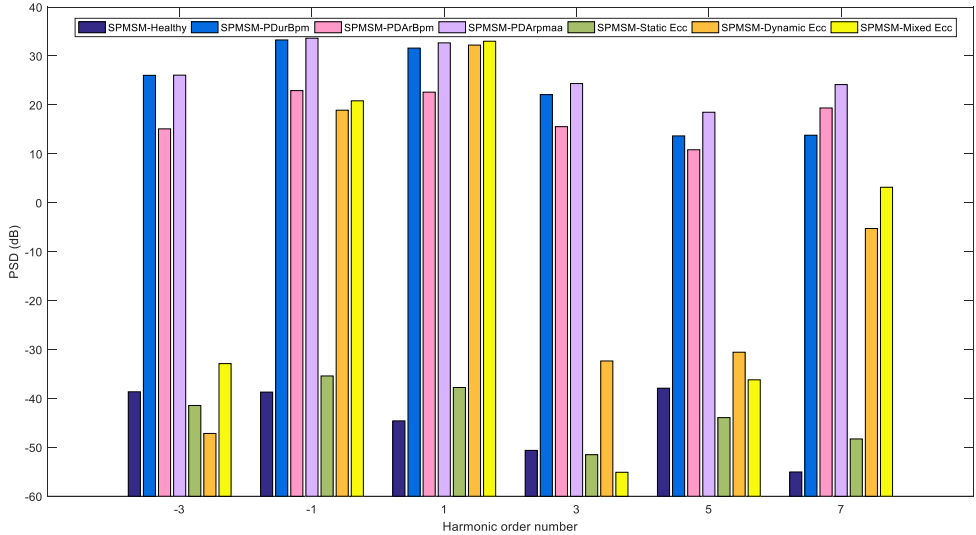

Fig. 8. Electromotive force harmonic components in decibels under healthy, faulty SPMSM (partial demagnetization and eccentricity faults)

## 5. Conclusion and future directions

Analysis and diagnosis of faults in SPMSM has been attracting a lot of attention in the automotive systemas it allows them to be more reliable and safe enhancing its sustainability. The first section was devoted to faults in PMSM drive train of HEV where we described electrical, mechanical and magnetic faults and their consequences on the perfermance of the propulsion system. In another major part of this paper, common fault diagnosis methods, are enumerated. This part demonstrates that research on analysis and diagnosis of faults is challenging and far from being completed. On the basis of analytical computation, two types of faults which are eccentricity and demagnetization faults of the SPMSM were studied analytically and by using FEM. The finites elements results are used to validate the analytical values of electromagnetic torque and the magnetic field. The results under different conditions obtained by simulation are presented and analyzed. As first results, we note that EMF magnitude varies depending on the width of the airgap and this variation is more significantly in case of ME. As second results, we note that the flux in cases partial demagnetization decreases compared to its value in healthy case. Likewise, the demagnetized PMs produces less torque than healthy PMs. Next, the signal processing methods is used as diagnosis method. So, this work mainly includes fault performance, harmonic component and signal spectrum characteristics, time-frequency analysis technique. In this study, we note that frequencies (1+3/P)fs and (1+5/P)fs are respectively more related to DE and SE. The frequency (1-3/P)fs is more related to ME. Too, the most dangerous fault is mixed eccentricity compared to static and dynamic eccentricity which reduces the machine performance.

In case of partial demagnetization, frequency (1+7/p) fs is more significative to detect the partial demagnetization types. Too, the most risky fault is the partial demagnetization fault by asymmetric reduction of a single PM arc angle compared to other demagnetization faults which reduces the machine performance.

Finally, study eccentricity fault experimentally can be a future research to validate the simulation results. Also, the harmonics amplitude will vary based on the motor operating conditions. Therefore, a study of the harmonics amplitude variation under different operating conditions and faults might be useful. Moreover, the data driven faults diagnosis methods is a relatively new research area, thus, several research challenges still remain open. Also estimating the fault severity, the post-fault life of the machine and ultimately fault tolerant control can be a future research area.